\documentclass[letters,fleqn,usenatbib]{mnras}

\usepackage[T1]{fontenc}
\usepackage{ae,aecompl}
\usepackage{adjustbox}
\usepackage{chngpage}

\usepackage{multicol}        
\usepackage{multirow}        

\usepackage{epsfig,amstext,floatfig,alltt,fancyhdr,setspace}
\usepackage{graphicx}
\usepackage{lmodern} 
\usepackage{color}
\usepackage{ucs}
\usepackage[utf8x]{inputenc}
\usepackage{url}
\usepackage{mathrsfs}
\usepackage{amsmath} 
\usepackage{amssymb}
\usepackage{appendix}
\usepackage{rjwmath}
\usepackage{varioref}
\usepackage{times}
\usepackage{enumerate}
\usepackage{magazin_abbrev}

\def \sect{Section\thinspace}
\def \fig{Fig.\thinspace}
\def \tab{Table\thinspace}

\def\P{{\cal P}}

\def\d{{\rm d}}

\newcommand{\mcal}{{\sc metacalibration }}
\newcommand{\im}{{\sc im3shape }}
\newcommand{\Om}{{\Omega_{\rm m}}}

\newcommand{\trox}{{T18 }}
\newcommand{\troxnospace}{{T18}}
\newcommand{\sysnospace}{{A18}}
\newcommand{\sys}{{A18 }}

\title[]
{Dark Energy Survey Year 1: An independent E/B-mode cosmic shear analysis}
\author[M. Asgari \& C. Heymans]{
Marika Asgari,$^{1}$\thanks{E-mail: ma@roe.ac.uk}
Catherine Heymans$^{1}$
\\
$^{1}$Institute for Astronomy, University of Edinburgh, Royal Observatory,
Blackford Hill, Edinburgh, EH9 3HJ, U.K.\\
}

\date{Accepted XXX. Received YYY; in original form ZZZ}

\pubyear{2018}

\begin{document}

\label{firstpage}
\pagerange{\pageref{firstpage}--\pageref{lastpage}}
\maketitle

\begin{abstract}
We present an independent cosmic shear analysis of the non-cosmological B-mode distortions within the public first year data from the Dark Energy Survey (DES).  We find no significant detection of B-modes in a full tomographic analysis of the primary \mcal shear catalogue.    This is in contrast to the secondary \im shear catalogue, where we detect B-modes at a significance of $\sim 3\sigma$ with a pattern that is consistent with the B-mode signature of a repeating additive shear bias across the survey.   We use the COSEBIs statistic to cleanly separate the B-modes from the gravitational lensing signal (E-modes).  We find good agreement between the measured E-modes and their theoretical expectation given the DES cosmological parameter constraints.
\end{abstract}

\begin{keywords}
Gravitational lensing: weak
\end{keywords}


\section{Introduction}
\label{sec:intro}

Cosmic shear is one of the primary cosmological probes of the large scale structures in the Universe \citep{BartelmannSchneider01}, with the potential to expand our understanding of dark matter and the accelerated expansion of the Universe. In order to release this potential we need to understand and control systematic effects that can contaminate this signal. In \citet[\sys hereafter]{asgari/etal:2018}, we extracted the systematic signal from three public cosmic shear surveys, the first 450 deg$^2$ from the Kilo Degree Survey \citep[KiDS-450,][]{hildebrandt/etal:2017}, the Canada France Hawaii Telescope Lensing Survey \citep[CFHTLenS,][]{Heymans12} and the Dark Energy Survey Science Verification \citep[DES-SV,][]{DES2016}. We used COSEBIs \citep[Complete Orthogonal Sets of E/B-Integrals,][]{SEK10}, which are two point statistics that cleanly separate, E-modes produced by the gravitational lensing and systematic effects from B-modes that originate from systematics only\footnote{Source clustering \citep{Schneider02} and higher order lensing effects \cite{Schneider98} can produce currently insignificant levels of B-modes. Some of the non-standard cosmological models are also able to produce them \citep[see for example][]{thomas/etal:2017}.}. 
By modelling a number of terrestrial systematic effects, we showed that the choice of the statistics used for E/B-mode decomposition is important and the systematics can affect E and B-modes at different scales. The signature of the systematic effects that we modelled can be compared to the data to diagnose the origin of the B-modes. 

In \sys we found evidence for systematics with a significant detection of B-mode for KiDS-450, CFHTLenS and DES-SV. We concluded the B-modes of all three surveys were consistent with a repeating additive shear bias systematic and that DES-SV, in addition, suffered from PSF-leakage and a redshift dependant selection bias. With the public release of DES first year data \citep[DES-Y1]{DES_DataRelease1} we revisit this analysis. \cite{Zuntz/etal:2018} presented a shear power spectrum analysis of DES-Y1 concluding  that the B-mode power was consistent with zero. As the main cosmological analysis of this dataset \citep[][\trox hereafter]{DES_y1_3x2pt, troxel/etal:2017} was carried out using the shear correlation functions, which are sensitive to different scales, we argue that this B-mode analysis is neither sufficient nor the appropriate method to then conclude zero B-mode contamination for the cosmological analysis that followed. We note that the same method was used for DES-SV which also concluded that the B-modes were insignificant in contrast to our analysis in \sys. 

To quantify the effect of the systematics on the cosmological parameters, \sys employed the data compression method of \cite{AS2014}, to produce compressed COSEBIs (CCOSEBIs). This method compresses the data according to its sensitivity to the parameters to be estimated. 
We compared the value of the measured CCOSEBIs on mock data with and without added systematics and concluded that all of the systematics that we had modelled added power to the signal.  As a result the measured values of $\sigma_8$ and $\Om$ are biased high when these systematics are present in the data and thus these systematics are unable to explain the persistent tension seen between cosmological parameter constraints from weak gravitational lensing surveys and those from the cosmic microwave background  \citep{hildebrandt/etal:2017,troxel/etal:2017, Hikage18, Planck2018}.

In this letter we measure COSEBIs and CCOSEBIs for DES-Y1 and compare them to the systematics modelled in \sysnospace.
COSEBIs and CCOSEBIs are introduced in \sect\ref{sec:Methods} and the data in \sect\ref{sec:Data}. The results are shown in \sect\ref{sec:Results} and the conclusions are laid out in \sect\ref{sec:Conclusions}.  Our pipeline to calculate COSEBIs is available on request.

\section{Methods}
\label{sec:Methods}

We use COSEBIs and CCOSEBIs to separate E/B-modes in DES-Y1. Here we only briefly outline the relevant equations and refer the reader to \cite{SEK10}, \cite{AS2014}  and \sys where they are introduced in more detail. 

\subsection{COSEBIs}

COSEBIs modes can be measured from survey data using their relation to shear two point correlation functions, $\xi_\pm$,
\begin{align}
\label{eq:EnReal}
 E_n &= \frac{1}{2} \int_{\theta_{\rm min}}^{\theta_{\rm max}}
 \d\theta\,\theta\: 
 [T_{+n}(\theta)\,\xi_+(\theta) +
 T_{-n}(\theta)\,\xi_-(\theta)]\;, \\
 \label{eq:BnReal}
 B_n &= \frac{1}{2} \int_{\theta_{\rm min}}^{\theta_{\rm
     max}}\d\theta\,\theta\: 
 [T_{+n}(\theta)\,\xi_+(\theta) -
 T_{-n}(\theta)\,\xi_-(\theta)]\;,
\end{align} 
where $E_n$ and $B_n$ are E and B-modes, respectively and $n$ is a natural number, making the COSEBIs modes discrete and well-defined. The filter functions, $T_{\pm n}(\theta)$, are defined such that $E_n$ and $B_n$ are pure E and B-modes without a trace of ambiguous modes. As $T_{\pm n}(\theta)$ produce a complete set of basis functions, COSEBIs modes are also complete. For cosmological inference we only expect to get a signal from $E_n$ for the first handful of n-modes \citep{Asgari12}. This is because $\xi_\pm$ are rather smooth and featureless functions and hence their behaviour can be characterised by a small number of modes. The $B_n$ can be used to track the remaining systematic effects, which we measure up to $n=20$ in this analysis.

\subsection{CCOSEBIs}

We use the data compression method defined in \cite{AS2014} to compress COSEBIs modes to a smaller number of quantities. This compression is very useful when tomographic data is considered, especially if the analysis combines cross-correlated data from different probes, as is the current trend in cosmology. 
\cite{AS2014} data compression relies on the sensitivity of the data vector to the model parameters. This sensitivity is quantified by the first and second order derivatives of the observable statistics to the parameters, as well as the covariance matrix of the statistics. The addition of the second order derivatives ensures that this data compression is less prone to losing information if the covariance matrix or the derivatives are not accurate. We call the compressed COSEBIs, CCOSEBIs, which are defined as,
\begin{equation}
\label{eq:CCOSEBIs}
\boldsymbol{E}^{\rm c}=\boldsymbol{\Gamma}\boldsymbol{E}~~~~~{\rm and} ~~~~~ \boldsymbol{B}^c=\boldsymbol{\Gamma}\boldsymbol{B}\;,
\end{equation}   
where $\boldsymbol{\Gamma}$ is the compression matrix, $\boldsymbol{E}^{\rm c}$ and $\boldsymbol{B}^{\rm c}$ are the E and B-mode CCOSEBIs, while $\boldsymbol{E}$ and $\boldsymbol{B}$ are the COSEBIs E and B-mode vectors. For $n_{\rm p}$ free parameters in the model and $N_{\rm tot}$ COSEBIs modes, $\boldsymbol{\Gamma}$  is an $n_{\rm p}(n_{\rm p}+3)/2 \times N_{\rm tot}$ matrix. The first $n_{\rm p}$ rows of $\boldsymbol{\Gamma}$ consist of the first order derivatives of $E_n$ with respect to the parameters multiplied by their inverse covariance matrix, while the rest of the rows correspond to the second order compressed modes, formed of the second order derivatives of $E_n$ multiplied by their inverse covariance matrix. 

The cosmic shear cosmological information is contained in the E-modes and the first few COSEBIs modes capture this information.  Therefore, the CCOSEBIs modes are formed of these modes. Here we only consider $\sigma_8$ and $\Om$ as free parameters and form the CCOSEBIs for these parameters, resulting in five CCOSEBIs modes, two first order modes: $E^{\rm c}_{\sigma_8}$ and $E^{\rm c}_{\Om}$ and three second order modes: $E^{\rm c}_{\sigma_8\sigma_8}$, $E^{\rm c}_{\sigma_8\Om}$ and $E^{\rm c}_{\Om\Om}$. This reduction in data volume can be compared to the 200 COSEBIs modes considered for a 4 tomographic bin analysis with 10 redshift bin pair combinations and $n=20$ mode measurements for each.

\section{Data Analysis}
\label{sec:Data}

The  first year data release of the Dark Energy Survey\footnote{https://des.ncsa.illinois.edu/releases/y1a1} 
\citep[DES-Y1,][]{DES_DataRelease1} consists of 1514 deg$^2$ of cosmic shear data in four photometric bands $griz$ \citep[see][for the image processing pipeline and the camera specifications]{morganson/etal:2018,flaugher/etal:2015}. 
This data has been analysed with two distinct shape measurement methods: \im  and \mcal \citep{Zuntz/etal:2018}. 
\mcal is a method that calibrates the results of any shape measurement method based on its sensitivity to shearing the image in different directions. For DES-Y1, \mcal is applied to shape measurements from a Gaussian model that is fit to the light profile of the galaxies as imaged in the $riz$ photometric bands, resulting in ellipticity measurements for $\sim$ 31 million galaxies. This is the primary catalogue for the cosmic shear analysis presented in \trox and \cite{DES_y1_3x2pt}, both of which use a subset of this catalogue that overlaps with their {\sc redMaGiC} galaxy sample, resulting in a reduced area of 1321 deg$^2$.
\im is a model fitting method that uses image simulations for calibration \citep{samuroff/etal:2018}. For the  DES-Y1 analysis, \im in contrast to \mcal analyses the $r$-band only, yielding a lower signal-to-noise catalogue with $\sim$ 21 million galaxies. This catalogue was primarily used for consistency checks. Here we show results for both catalogues for the full available area.   

\trox measured correlation functions between $\theta\in[2.5',250']$, but used a variable minimum angular distance for their cosmological analysis, to avoid sensitivity to baryonic feedback \citep{Semboloni11}. They used either $\theta_{\rm min}=4'$ or $\theta_{\rm min}=8'$ for $\xi_+$ and $\theta_{\rm min}=40'$ or $\theta_{\rm min}=100'$ for $\xi_-$, keeping $\theta_{\rm max}=250'$ in all cases. In this paper we analyse these five angular ranges: $[0.5',250']$, $[4',250']$, $[8',250']$, $[40',250']$ and $[100',250']$, which cover all ranges analysed by \trox with the addition of a larger angular range, $[0.5',250']$, anticipating future analyses that may use smaller scales \citep[see for example][]{maccrann/etal:2017}. 
Following  \trox we divide the data into four photometric-redshift bins: $z_{\rm phot}\in(0.2 , 0.43)$,  $z_{\rm phot}\in(0.43 ,0.63)$,  $z_{\rm phot}\in(0.63 , 0.9)$,  $z_{\rm phot}\in(0.9 , 1.3)$ and also include a combined bin of $z_{\rm phot}\in(0.2 , 1.3)$. We show predictions for the E-modes using the best fitting cosmological and nuisance parameters assuming a flat-$\Lambda$CDM model based on the cosmic shear only analysis of \trox (see \tab\ref{tab:CosmoParam}) and the redshift distributions described in \cite{Hoyle/etal:2018}. The theory predictions are calculated with {\sc cosmosis} \citep{cosmosis}\footnote{{\sc cosmosis}: bitbucket.org/joezuntz/cosmosis} using the setup adopted by \trox: {\sc camb} for the linear matter power spectrum \citep{camb2000,camb12}\footnote{{\sc camb}: http://camb.info}, \cite{Takahashi12} for the nonlinear evolution of matter, \cite{Bridle07}\footnote{\texttt{bk\char`_corrected} in {\sc cosmosis}} for the intrinsic alignments with an additional redshift dependence characterised by a power and a pivot redshift (see \tab\ref{tab:CosmoParam}). In addition, we include the four multiplicative shear calibration parameters as well as another four additive photometric redshift bias parameters in our predictions. For the single bin theory predictions we set the nuisance parameters to zero and find that this provides a good fit to the non-tomographic data. 


To calculate COSEBIs we measure 2PCFs using {\sc athena}\footnote{{\sc athena}: www.cosmostat.org/software/athena} \citep{KilbingerAthena14} with a million linear angular bins in $[0.5', 250']$. See \cite{asgari/etal:2017} for a series of accuracy tests when measuring COSEBIs.

\begin{table}
\centering
\caption{\small{The best-fitting cosmological parameters for DES-Y1 \mcal (Mcal) and \im (Im3) catalogues (T18), estimated from their cosmological chains. $\sigma_8$ is the standard deviation of perturbations in a sphere of radius $8 h^{-1} {\rm Mpc}$, today.  
$n_{\rm s}$ is the spectral index of the primordial power spectrum. $\Om$, $\Omega_{\rm b}$ and  $\Omega_\nu$ are the matter,
 baryon and neutrino density parameters, respectively and $h$ is the dimensionless Hubble parameter. 
The underlying cosmology is a flat-$\Lambda$CDM model with Gaussian initial perturbations. 
The last two columns show the intrinsic alignment parameters: $A_{\rm IA}$, 
the amplitude of the intrinsic galaxy alignment model and $\eta$ 
characterising the redshift dependence of the intrinsic alignments around a pivot redshift of $z_0=0.62$. }}
\label{tab:CosmoParam}
\begin{adjustwidth}{-6mm}{0mm}
\resizebox{90mm}{!}{
\begin{tabular}{ c  c  c  c  c  c  c  c  c }
  & $\sigma_8$ & $\Om$   & $n_{\rm s}$  & $h$    & $\Omega_{\rm b}$ & $\Omega_\nu$ & $A_{\rm IA}$ & $\eta$ \\
\hline
Mcal & 0.81              & 0.29       & 0.97              &  0.74   &   0.033                  & 0.010              &  1.40             & 1.43    \\ \hline
Im3 & 0.63              & 0.48       & 0.92              & 0.62    &   0.031                  & 0.016              &  0.48             &  -0.41  
\end{tabular}
}
\end{adjustwidth}
\end{table}

\section{Results}
\label{sec:Results}

We present COSEBIs and CCOSEBIs measurements for both \im and \mcal catalogues, focusing on the significance of the measured B-modes. To do so we measure $p$-values for the B-modes given a null model with a Gaussian distribution. A $p$-value shows the probability of measuring a B-mode signal that is equally or more extreme than the one measured from the data, given the model (see Appendix C of \sys for more details). To calculate the $p$-values we need to first calculate the $\chi^2$ values for which a covariance matrix is needed. In the absence of systematics or physical phenomena that can produce B-modes, this covariance matrix will only depend on the weighted number of galaxy pairs, $N_{\rm pair}$, and the weighted dispersion in galaxy shapes, $\sigma_\epsilon$ \footnote{This analytical result is validated against simulations in \cite{asgari/etal:2017}.}. We use the measured $N_{\rm pair}$ and $\sigma_\epsilon$ from the data to form the covariance matrix. When comparing the E-modes to their expectation values based on the \trox  best-fit parameters in \tab\ref{tab:CosmoParam}, we add the cosmic variance terms to the Gaussian covariance but ignore the sub-dominant non-Gaussian and super sample terms. 

The covariance matrix for the CCOSEBIs can be derived from the COSEBIs covariance, by multiplying both sides with the compression matrix, $\boldsymbol{\Gamma}$. More details about the covariance calculation is given in the method section of \sysnospace. 

\begin{figure*}
   \begin{center}
     \begin{tabular}{c}
     \resizebox{160mm}{!}{\includegraphics{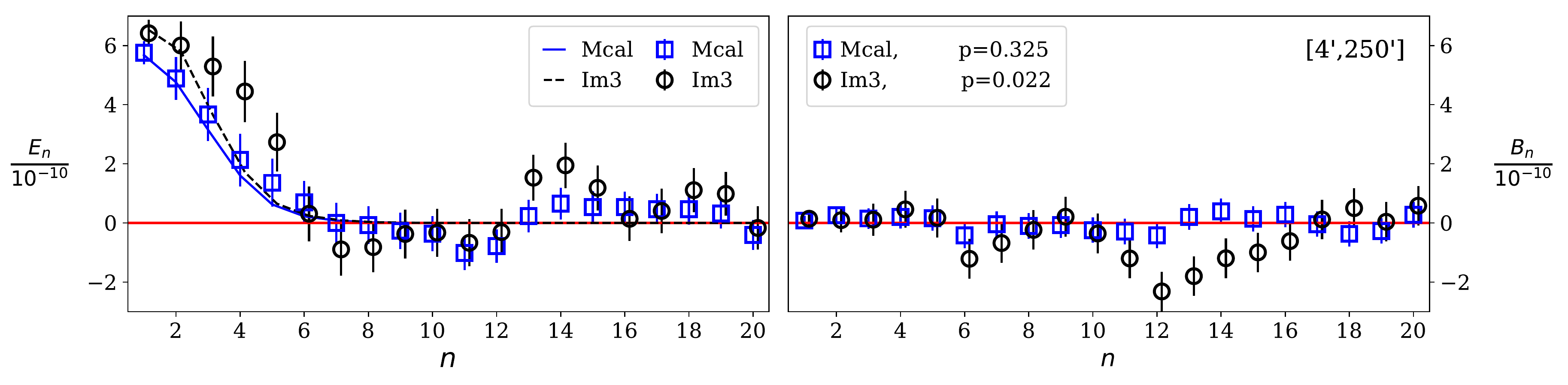}}
     \end{tabular}
   \end{center}
     \caption{\small{COSEBIs E-modes (left) and B-modes (right) for DES-Y1 catalogues. \mcal results are shown in blue and \im in black. The $p$-values shown in the legends correspond to the B-modes assuming a null hypothesis of zero B-modes. The angular range considered here is $[4',250']$. Note that COSEBIs modes are discrete and the theory values are connected to each other only for visual aid. The errors for E-modes include cosmic variance terms and hence are larger than the B-mode errorbars.}}
     \label{fig:COSEBIs1bin}
 \end{figure*}

We first look at the COSEBIs measurements for a non-tomographic analysis in \fig\ref{fig:COSEBIs1bin} showing results for \mcal (blue squares) and \im (black circles). Here we only show results for the $[4',250']$ angular range, that corresponds to the widest angular range used in \troxnospace. On the left hand side we show E-modes and on the right B-modes finding the B-modes for \mcal to be consistent with zero. In contrast, the \im E/B-modes show unexpected oscillations that resemble the systematic signature of a repeating additive shear bias (see figure 10 in \sysnospace). This feature was also apparent in the DES-SV COSEBIs analysis (\sysnospace).

\begin{figure*}
   \begin{center}
     \begin{tabular}{c}
     \resizebox{83mm}{!}{\includegraphics{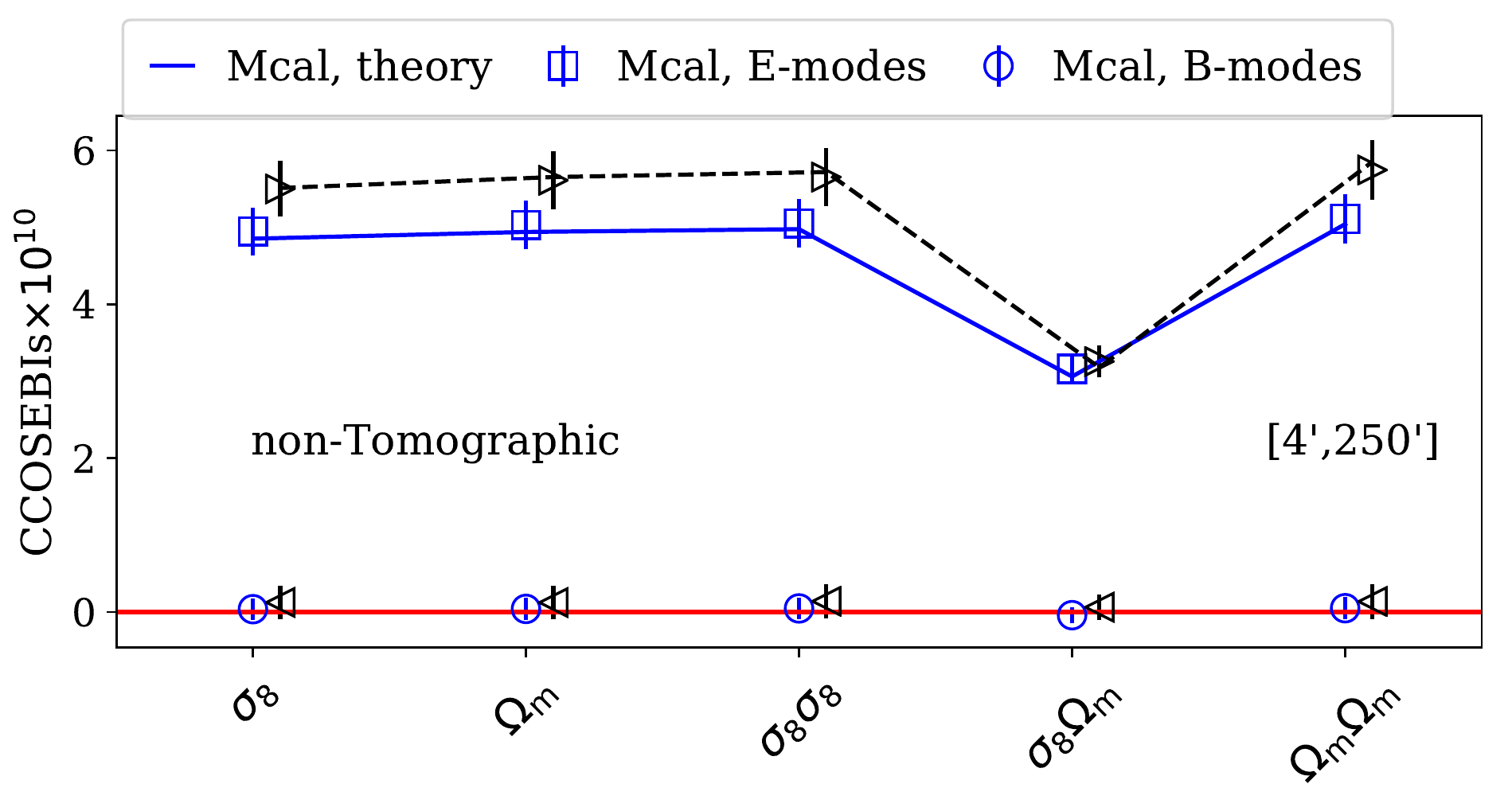}}
     \resizebox{83mm}{!}{\includegraphics{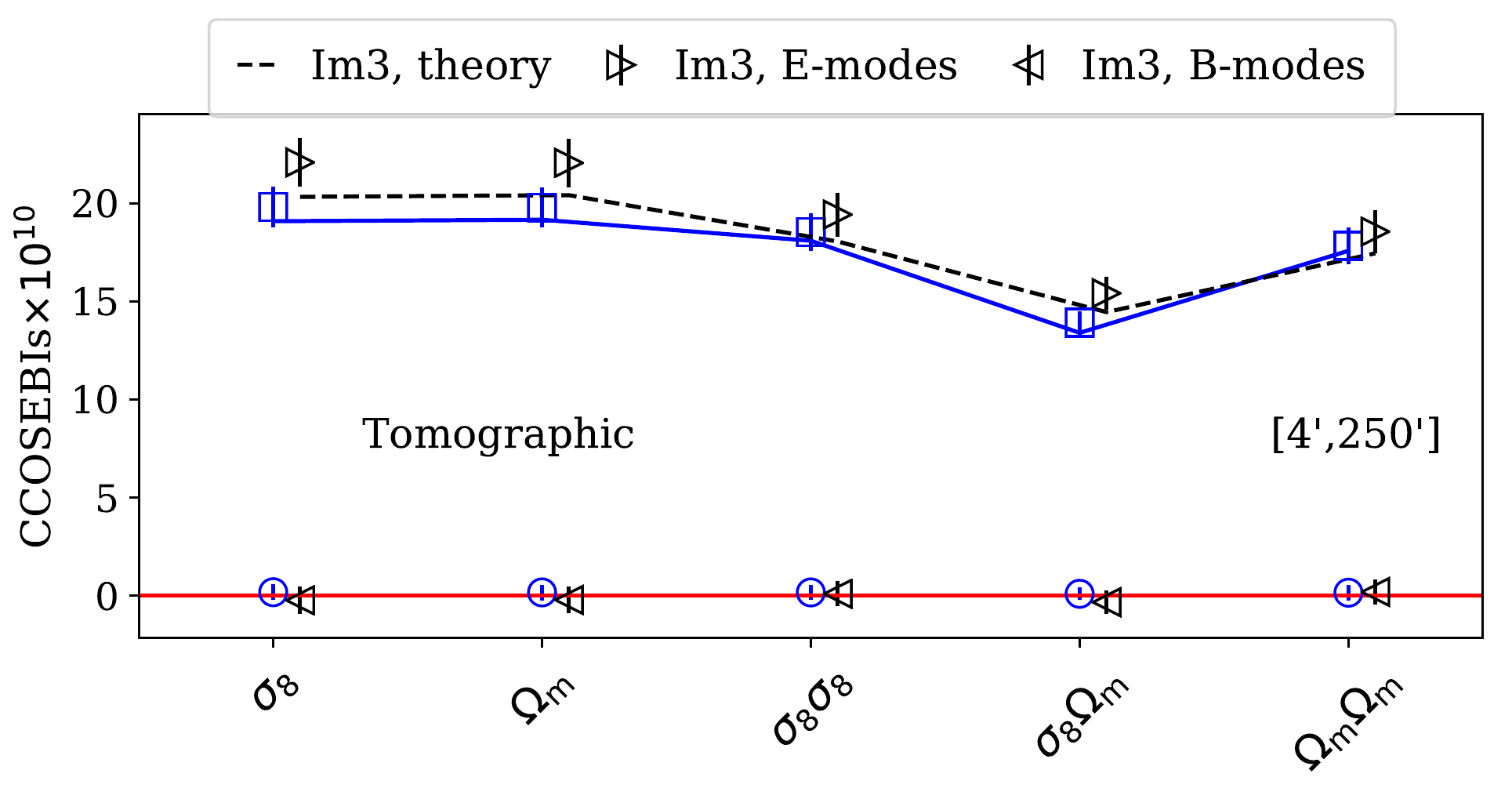}}
     \end{tabular}
   \end{center}
     \caption{\small{Compressed COSEBIs results for \mcal (blue) and \im (black) catalogues. The angular separation range is $[4',250']$ and the results are shown for non-tomographic (left) and tomographic (right) cases. B-modes are shown to be consistent with zero for both catalogues. The expected theory values for E-modes are shown as curves (see \tab\ref{tab:CosmoParam} for the input cosmological parameters). Note that CCOSEBIs modes are discrete and the theory values are connected to each other only for visual aid. }}
\label{fig:CCOSEBIs}
\end{figure*}

In \fig\ref{fig:CCOSEBIs} we present the CCOSEBIs analysis for the non-tomographic (left) and tomographic (right) cases. The blue symbols and curves represent \mcal (squares: E-modes and circles: B-modes) and the black symbols and the dashed curve represent \im (right facing triangles: E-modes and left facing triangles: B-modes). The angular range used here is the same as in \fig\ref{fig:COSEBIs1bin}, with $\theta\in[4',250']$. The E-modes match their predicted theory values well. CCOSEBIs for $\sigma_8$ and $\Om$ are highly correlated and we caution the reader if carrying out a visual inspection.  

We summarise our analysis for the other angular ranges in the form of $p$-values for the B-modes in Tables \ref{tab:pvaluemcal}, \ref{tab:pvalueim3}, \ref{tab:pvaluemcal_tomo} and \ref{tab:pvalueim3_tomo}. All  $p$-values$\;<0.01$ are shown as bold ($>2.3\sigma$ detection). Tables \ref{tab:pvaluemcal} and \ref{tab:pvalueim3}  show $p$-values for the COSEBIs and CCOSEBIs B-mode measurements for 5 angular ranges and for the single bin and tomographic cases. In \tab\ref{tab:pvaluemcal} we show results for {\sc metacalibration}, while in \tab\ref{tab:pvalueim3} \im results are shown. We see that the $p$-values show no significant B-modes for the primary \mcal DES-Y1 shape catalogues, for all angular ranges. This is, however, not the case for the \im results, where, there is a detection of COSEBIs B-modes in the $[0.5',250']$ angular range for both tomographic ($3.4\sigma$) and non-tomographic ($2.3\sigma$) cases as well as the tomographic case for $[4',250']$ ($2.8\sigma$).  The CCOSEBIs results which are not sensitive to higher COSEBIs modes do not show any significant B-modes, however as we have shown in \sysnospace, this does not guarantee that the E-modes are not affected by these systematics, since some systematic effects, such as a repeating bias pattern have a different signature for E and B-modes. For these systematics even though they affect higher order B-modes more significantly, they tend to produce E-modes for the lower COSEBIs modes. As a result they will bias the cosmological analysis.

\begin{table}
\centering
\caption{\small{B-mode $p$-values for the \mcal catalogue. Each row shows a different angular separation range. The first two columns of numbers show $p$-values for COSEBIs with 20 modes and the last two columns belong to compressed COSEBIs. All $p$-values smaller than 0.01, corresponding to 2.3 $\sigma$, are shown in boldface. The results are shown for both tomographic and non-tomographic cases. }}
\label{tab:pvaluemcal}
\begin{adjustwidth}{0mm}{0mm}
\noindent\resizebox{80mm}{!}{
\begin{tabular}{ l c c  c  c  c }
 & \multicolumn{2}{c}{COSEBIs}&\multicolumn{2}{c}{CCOSEBIs}
 \\ \hline
$\theta$ range & Single Bin & Tomo\;\;\;\;   & Single Bin & Tomo \\ \hline
$[0.5',250']$ &$ 0.16$ &$ 0.11$ & $ 0.20$ & $ 0.07$
 \\ \cline{1-5}
$[4',250']$ &$ 0.33$ &$ 0.35$ & $ 0.88$ & $ 0.96$
 \\ \cline{1-5}
$[8',250']$ &$ 0.67$ &$ 0.10$ & $ 0.34$ & $ 0.64$
 \\ \cline{1-5}
$[40',250']$ &$ 0.80$ &$ 0.17$ & $ 0.91$ & $ 0.78$
 \\ \cline{1-5}
$[100',250']$ &$ 0.46$ &$ 0.03$ & $ 0.76$ & $ 0.25$
 \\ \cline{1-5}
\end{tabular}
}
\end{adjustwidth}
\end{table} 
\begin{table}
\caption{\small{Same as \tab\ref{tab:pvaluemcal}, but for \im catalogues.}}
\label{tab:pvalueim3}
\begin{adjustwidth}{0mm}{0mm}
\noindent\resizebox{80mm}{!}{
\begin{tabular}{ l  c  c  c  c }
 & \multicolumn{2}{c}{COSEBIs}&\multicolumn{2}{c}{CCOSEBIs}
 \\ \hline
$\theta$ range & Single Bin & Tomo\;\;\;\; & Single Bin & Tomo \\ \hline
$[0.5',250']$ &$\boldsymbol{9.8e-3}$ & $\boldsymbol{3.7e-4}$ & $ 0.23$ & $ 0.07$
 \\ \cline{1-5}
$[4',250']$ &$ 0.02$ & $\boldsymbol{2.2e-3}$ & $ 0.89$ & $ 0.44$
 \\ \cline{1-5}
$[8',250']$ &$ 0.06$ &$ 0.01$ & $ 0.17$ & $ 0.22$
 \\ \cline{1-5}
$[40',250']$ &$ 0.10$ &$ 0.06$ & $ 0.30$ & $ 0.89$
 \\ \cline{1-5}
$[100',250']$ &$ 0.11$ &$ 0.03$ & $ 0.19$ & $ 0.87$
 \\ \cline{1-5}
\end{tabular}
}
\end{adjustwidth}
\end{table} 
$\chi^2$ is a quantity that compresses all the information into a single summary statistic and takes a weighted average over the data, therefore, it is not always the most informative quantity. As a result, in Tables \ref{tab:pvaluemcal_tomo} and \ref{tab:pvalueim3_tomo} we separately analyse the auto and cross-correlations in the tomographic data for the different redshift bin pairs and calculate $p$-values for COSEBIs in each case. \tab\ref{tab:pvaluemcal_tomo} shows results for \mcal and \tab\ref{tab:pvalueim3_tomo} for {\sc im3shape}. Here we see that the significant B-modes detected for \im do not result in a significant $p$-value for any of the redshift bin pairs individually, but as many of these $p$-values are close to the threshold, their "averaged" $p$-value becomes significant. On the other hand we see that although none of the \mcal $p$-values were significant in \tab\ref{tab:pvaluemcal}, the combination of the two highest redshift bins, $z$-34,  shows significant B-modes of up to $3.0\sigma$ for 3 of the angular ranges. On inspection, the B-mode in the $z$-34 bin combination has oscillatory features out to high $n$-modes, similar to the pattern expected from a repeating additive shear bias.  When all the bins are combined, however, this significant detection is cancelled by the large $p$-values for the lower redshift bins. 
The distribution of the \mcal $p$-values in \tab\ref{tab:pvaluemcal_tomo} is essentially flat which is expected if the model (zero B-modes) is a good description of the data.  The distribution of the \im $p$-values in \tab\ref{tab:pvalueim3_tomo}, however, shows a visible skewness towards smaller values which indicates that the zero B-mode model is a poor fit (see Appendix C in A18). 
\begin{table}
\centering
\caption{\small{B-mode $p$-values for \mcal catalogues split for each pair of redshift bins. Each row presents $p$-values for the given redshift bin pair, z-$ij$, corresponding to the COSEBIs measured for redshift bins $i$ and $j$. The results are shown for 5 angular ranges and 20 COSEBIs modes. The boldfaced values correspond to $p$-value\;<\;0.01 and hence a significant B-mode detection, larger than  2.3 $\sigma$.}}
\label{tab:pvaluemcal_tomo}
\begin{adjustwidth}{-5mm}{0mm}
\noindent\resizebox{87mm}{!}{
\begin{tabular}{ c  c  c  c  c  c }
& $[0.5'$, $250']$    & $[4'$, $250']$    & $[8'$, $250']$    & $[40'$, $250']$    & $[100'$, $250']$     \\ \hline
z-11    &$ 1.00$ &$ 1.00$ &$ 0.99$ &$ 0.86$ &$ 0.60$
 \\ \hline
z-12    &$ 0.84$ &$ 0.90$ &$ 0.59$ &$ 0.80$ &$ 0.68$
 \\ \hline
z-13    &$ 0.49$ &$ 0.41$ &$ 0.67$ &$ 0.43$ &$ 0.69$
 \\ \hline
z-14    &$ 0.48$ &$ 0.35$ &$ 0.77$ &$ 0.60$ &$ 0.75$
 \\ \hline
z-22    &$ 0.04$ &$ 0.12$ &$ 0.11$ &$ 0.85$ &$ 0.38$
 \\ \hline
z-23    &$ 0.18$ &$ 0.63$ &$ 0.48$ &$ 0.95$ &$ 0.52$
 \\ \hline
z-24    &$ 0.13$ &$ 0.07$ &$ 0.03$ &$ 0.03$ &$ 0.22$
 \\ \hline
z-33    &$ 0.06$ &$ 0.41$ &$ 0.16$ &$ 0.11$ &$ 0.15$
 \\ \hline
z-34    &$ 0.09$ &$ 0.03$ & $\boldsymbol{1.2e-3}$ & $\boldsymbol{2.1e-3}$ & $\boldsymbol{1.6e-3}$
 \\ \hline
z-44    &$ 0.51$ &$ 0.71$ &$ 0.79$ &$ 0.22$ &$ 0.01$
 \\ \hline
\end{tabular}
}
\end{adjustwidth}
\end{table}
\begin{table}
\centering
\caption{\small{Same as \tab\ref{tab:pvaluemcal_tomo} but for \im catalogues.}}
\label{tab:pvalueim3_tomo}
\begin{adjustwidth}{-5mm}{0mm}
\noindent\resizebox{87mm}{!}{
\begin{tabular}{ c  c  c  c  c  c }
& $[0.5'$, $250']$    & $[4'$, $250']$    & $[8$', $250']$    & $[40'$, $250']$    & $[100'$, $250']$     \\ \hline
z-11    &$ 0.03$ &$ 0.03$ &$ 0.04$ &$ 0.54$ &$ 0.35$
 \\ \hline
z-12    &$ 0.03$ &$ 0.03$ &$ 0.29$ &$ 0.16$ &$ 0.04$
 \\ \hline
z-13    &$ 0.43$ &$ 0.92$ &$ 0.80$ &$ 0.58$ &$ 0.21$
 \\ \hline
z-14    &$ 0.11$ &$ 0.02$ &$ 0.06$ &$ 0.05$ &$ 0.03$
 \\ \hline
z-22    &$ 0.29$ &$ 0.47$ &$ 0.43$ &$ 0.60$ &$ 0.36$
 \\ \hline
z-23    &$ 0.19$ &$ 0.49$ &$ 0.26$ &$ 0.79$ &$ 0.89$
 \\ \hline
z-24    &$ 0.25$ &$ 0.25$ &$ 0.23$ &$ 0.05$ &$ 0.28$
 \\ \hline
z-33    &$ 0.10$ &$ 0.08$ &$ 0.05$ &$ 0.33$ &$ 0.49$
 \\ \hline
z-34    &$ 0.20$ &$ 0.10$ &$ 0.34$ &$ 0.21$ &$ 0.22$
 \\ \hline
z-44    &$ 0.03$ &$ 0.17$ &$ 0.19$ &$ 0.16$ &$ 0.19$
 \\ \hline
\end{tabular}
}
\end{adjustwidth}
\end{table}

\section{Conclusions}
\label{sec:Conclusions}
One of the major challenges for current and future cosmic shear analysis is the appropriate treatment of systematic effects. With COSEBIs we can separate E/B-modes completely and efficiently. The B-modes of COSEBIs are very sensitive to the systematic effects in the data and can be used as a diagnosis tool. In addition, its E-modes can be used for cosmological analysis, allowing for a consistent treatment of the data for both E and B-modes. Here we followed the analysis of \cite{asgari/etal:2018} measuring COSEBIs and CCOSEBIs E/B-modes for both catalogues of the DES-Y1 data. We found that \mcal has B-modes that are consistent with zero for the full tomographic analysis.  On inspection of individual bin combinations, however, we found the $z$-34 bin to be an outlier with a $3.0\sigma$ B-mode.   This low-level B-mode from a single bin is unexpected to produce any biases in the cosmological analysis for the current volume of the data. For future data releases, however, this signal should be investigated further as the last redshift bins have the highest signal-to-noise contribution to the cosmological analysis. 

A tomographic analysis of the \im catalogue shows a significant detection of B-modes up to $3.4\sigma$.
Although \im is not used in the primary DES cosmological analysis it is used for validation. If the B-modes in the \im catalogues persist, their use as a validation tool becomes limited.

\cite{hildebrandt/etal:2018} presents an updated cosmic shear analysis of the Kilo-Degree Survey (KiDS) that incorporates near infra-red photometry \citep{wright/etal:2018} and a number of other improvements \citep[][]{kannawadi/etal:2018}.  They conclude that their control of systematics has significantly improved in this new 9-band analysis, with the previously reported significant B-mode detection found to now be consistent with zero. Given this development for KiDS, and our independent verification of the high quality systematic control within DES, we see a very promising future for high accuracy cosmic shear cosmology.

\vspace{-5mm}
\section*{Acknowledgements}
\small{We thank Ami Choi, Niall MacCrann, Michael Troxel  and Joe Zuntz for their help with using the DES-Y1 products and useful discussions. We thank Vasiliy Demchenko and the referee for useful comments. We acknowledge support from the European Research Council (grant number 647112).}

\footnotesize{This project used public archival data from the Dark Energy Survey (DES). Funding for the DES Projects has been provided by the U.S. Department of Energy, the U.S. National Science Foundation, the Ministry of Science and Education of Spain, the Science and Technology Facilities Council of the United Kingdom, the Higher Education Funding Council for England, the National Center for Supercomputing Applications at the University of Illinois at Urbana–Champaign, the Kavli Institute of Cosmological Physics at the University of Chicago, the Center for Cosmology and Astro-Particle Physics at the Ohio State University, the Mitchell Institute for Fundamental Physics and Astronomy at Texas A\&M University, Financiadora de Estudos e Projetos, Funda{\c c}{\~a}o Carlos Chagas Filho de Amparo {\`a} Pesquisa do Estado do Rio de Janeiro, Conselho Nacional de Desenvolvimento Cient{\'i}fico e Tecnol{\'o}gico and the Minist{\'e}rio da Ci{\^e}ncia, Tecnologia e Inova{\c c}{\~a}o, the Deutsche Forschungsgemeinschaft, and the Collaborating Institutions in the Dark Energy Survey.
The Collaborating Institutions are Argonne National Laboratory, the University of California at Santa Cruz, the University of Cambridge, Centro de Investigaciones Energ{\'e}ticas, Medioambientales y Tecnol{\'o}gicas-Madrid, the University of Chicago, University College London, the DES-Brazil Consortium, the University of Edinburgh, the Eidgen{\"o}ssische Technische Hochschule (ETH) Z{\"u}rich,  Fermi National Accelerator Laboratory, the University of Illinois at Urbana-Champaign, the Institut de Ci{\`e}ncies de l'Espai (IEEC/CSIC), the Institut de F{\'i}sica d'Altes Energies, Lawrence Berkeley National Laboratory, the Ludwig-Maximilians Universit{\"a}t M{\"u}nchen and the associated Excellence Cluster Universe, the University of Michigan, the National Optical Astronomy Observatory, the University of Nottingham, The Ohio State University, the OzDES Membership Consortium, the University of Pennsylvania, the University of Portsmouth, SLAC National Accelerator Laboratory, Stanford University, the University of Sussex, and Texas A\&M University.
Based in part on observations at Cerro Tololo Inter-American Observatory, National Optical Astronomy Observatory, which is operated by the Association of Universities for Research in Astronomy (AURA) under a cooperative agreement with the National Science Foundation.\vspace{-5mm}}
\bibliographystyle{mnras}
\bibliography{COSEBIs} 
\bsp	
\label{lastpage}
\end{document}